\documentclass{article}

\usepackage{arxiv}

\usepackage[utf8]{inputenc} 
\usepackage[T1]{fontenc}    
\usepackage{hyperref}       
\usepackage{url}            
\usepackage{booktabs}       
\usepackage{amsfonts}       
\usepackage{nicefrac}       
\usepackage{microtype}      
\usepackage{lipsum}
\usepackage{graphicx}
\usepackage{svg}
\usepackage{balance}  
\usepackage{xcolor}
\usepackage{amssymb,amsmath,amsthm}
\usepackage{hyperref}

\usepackage{amsmath}
\usepackage{bbm}
\usepackage{booktabs}
\usepackage{comment}
\usepackage{makecell}

\usepackage{float}
\usepackage{algorithmic}
\usepackage{algorithm, setspace}
\usepackage[english]{babel}
\usepackage{bm}
\usepackage{mathtools}
\usepackage{breqn}

\usepackage[font=small,labelfont=bf,tableposition=top]{caption}

\DeclareCaptionLabelFormat{andtable}{#1~#2  \&  \tablename~\thetable}

\newtheorem{theorem}{Theorem}[section]
\newtheorem{lemma}[theorem]{Lemma}

\title{RAMBO: Repeated And Merged BloOm Filter for Ultra-fast Multiple Set Membership Testing (MSMT) on Large-Scale Data}

\author{
Gaurav Gupta \\
  Rice University\\
  Houston, Texas, USA \\
  \texttt{gaurav.gupta@rice.edu} \\
  \And 
   Minghao Yan \\
  Rice University\\
  Houston, Texas, USA \\
  \texttt{my29@rice.edu} \\
  \And
   Benjamin Coleman \\
  Rice University\\
  Houston, Texas, USA \\
  \texttt{ben.coleman@rice.edu} \\
  \And
   R. A. Leo Elworth \\
  Rice University\\
  Houston, Texas, USA \\
  \texttt{chilleo@gmail.com} \\
  \And
  Tharun Medini \\
  Rice University\\
  Houston, Texas, USA \\
  \texttt{tharun.medini@rice.edu} \\
  \And
  Todd Treangen \\
  Rice University\\
  Houston, Texas, USA \\
  \texttt{treangen@rice.edu} \\
  \And
  Anshumali Shrivastava \\
  Rice University\\
  Houston, Texas, USA \\
  \texttt{anshumali@rice.edu} \\
}

\begin{document}

\maketitle

\begin{abstract}

Multiple Set Membership Testing (MSMT) is a well known problem in a variety of search and query applications. 
Given a dataset of $K$ different sets and a query $q$, it aims to find all of the sets containing the query. Trivially, an MSMT instance can be reduced to $K$ membership testing instances, each with the same $q$, leading to $O(K)$ query time with a simple array of Bloom Filters.
We propose a data-structure called RAMBO (Repeated And Merged BloOm Filter) that achieves $O(\sqrt{K}\log {K})$ query time in expectation with an additional worst-case memory cost factor of $O(\log K)$ beyond the array of Bloom Filters. Due to this, RAMBO is a very fast and accurate data-structure. 
Apart from being embarrassingly parallel, supporting cheap updates for streaming inputs, zero false-negative rate, and low false-positive rate, RAMBO beats the state-of-the-art approaches for genome indexing methods- COBS (Compact bit-sliced signature index), Sequence Bloom Trees (a Bloofi based implementation), HowDeSBT, SSBT, and document indexing methods like BitFunnel. The proposed data-structure is simply a count-min sketch type arrangement of a membership testing utility (Bloom Filter in our case). It indexes $k$-grams and provides an approximate membership testing based search utility. The simplicity of the algorithm and embarrassingly parallel architecture allows us to index a 170 TB genome dataset in a mere 14 hours on a cluster of 100 nodes while competing methods require weeks.

\end{abstract}

\section{Introduction}
Approximate set membership is a fundamental problem that arises in many high-speed memory-constrained applications in databases, networking, and search. The Bloom Filter~\cite{bloom1970space,mitzenmacher2002compressed,cohen2003spectral} is one of the most famous and widely adopted space-efficient data structures for approximate set membership. It allows constant time, i.e., $O(1)$ membership testing in mere $O(n)$ space, where $n$ is the cardinality of the set $S$ under consideration. Bloom Filters trade a small false-positive probability for an impressive reduction in query time and memory usage. Bloom Filters are successful in many latency and memory-constrained scenarios compared to sophisticated hashing algorithms~\cite{networkapplications,simpleMPH} because of their simplicity and ability to cheaply insert elements on the fly.

In this work, we focus on the multiple set membership testing (MSMT) problem. Here, instead of a single set $S$, we are given a set of $K$ sets $\mathcal{S} = \{ S_1, \ S_2, ..., \ S_K \}$. Each set $S_i$ contains elements, called ``keys'', from a universe $\Omega$ of all possible keys. Given a query $q \in \Omega$, our goal is to identify all the sets in $\mathcal{S}$ that contain $q$. That is, the task is to return the subset $A_q \subseteq \mathcal{S}$, such that $q \in S_i$ if and only if $S_i \in A_q$. 

\subsection{Related Works}
Indexing and search problems can be solved in the MSMT framework by treating documents as sets and terms as keys. Traditional approaches such as the inverted index cannot quickly index large-scale data without violating memory constraints. Bloom Filters~\cite{bloom1970space} and similar bit-signature approaches have been used extensively in recent years for indexing genomic data and web documents. The open source Bing search engine uses BitFunnel \cite{BitFunnel}, an index which uses document hash signatures in a Bloom Filter-like structure for fast search. Computational biologists have also shifted to Bloom Filter methods for gene sequence search due to the sheer scale of genomic data. Practical methods range from ranking rows - BitFunnel~\cite{BitFunnel}, tree based filter structures - Bloofi ~\cite{crainiceanu2013bloofi}, Sequence Bloom Trees (SBT), HowDeSBT \cite{HowDeSBT}, Split-SBT ~\cite{kodama2011sequence,b2}, to a simple array of $K$ Bloom Filters in BIGSI and ~\cite{b3} COBS~\cite{Bingmann2019COBSAC}. Refer Table \ref{tab:relatedwork}. Such methods have made a significant impact in taming the terabytes and petabytes of data in genomics and web search. RAMBO is the next attempt. It approaches the problem of indexing massive dataset, in a very intuitive way by creating merges and repetition of Bloom Filter based membership testing. It beats the current baselines by achieving a very robust, low memory and ultra-fast indexing data-structure. 

\begin{table}[H]
\centering
\fontsize{9}{14}\selectfont

\begin{tabular}{|c|c|c|c|}
\hline
Method & Memory  & Query & Comments \\
\hline
Inverted Index & $\log K \sum_{S\in\mathcal{S}}|S|$  & $O(1)$ & Takes huge time for creation \\
 & &  & Extra space for posting lists \\ 
\hline
BIGSI/COBS & $\sum_{S\in\mathcal{S}}|S|$ & $O(K)$ & \\ \hline
Sequence Bloom Trees & $\log K \sum_{S\in\mathcal{S}}|S| $ & Worst case: $O(K)$  & Sequential query in trees\\ 
 &  &  Best case: $O(logK)$  & \\
\hline
RAMBO  & $\Gamma \log K  \sum_{S \in \mathcal{S}} |S| $ & $O(\sqrt{K} \log K)$ & $\Gamma <1$ (data-dependent)\\
\hline
\end{tabular}
\caption{Theoretical comparison of related algorithms on MSMT problem. $S \in \mathcal{S}$ represents a set. $K$ is total number of sets. Refer Section \ref{TheroryAnalysis} for RAMBO.}
\label{tab:relatedwork}
\end{table}

\subsection{Our Contribution}
We introduce a new data structure for the MSMT problem that prioritizes query time. In particular, we propose a method with efficient query time and these additional properties: 1) Low false-positive rate 2) Zero false-negative rate 3) An embarrassingly parallel data structure and 4) Cheap updates for streaming inputs. With this, we show a remarkably improved capability, over the baselines, of constructing and querying with the 170TB WGS dataset \cite{b3}, a fractional wiki dump~\cite{wiki-dump} and ClubWeb09 data~\cite{clueweb}. We achieved between \textbf{25x} to \textbf{2000x} improvement in query time over the most competitive baseline of gene data indexing while keeping competitive false-positive rates. Although RAMBO uses slightly more indexing memory than the optimal array of Bloom Filter (COBS), it is capable to keep a cheap index (1.8 terabytes) for 170 terabytes worth of data. It is important to note that Bloom Filter in RAMBO can be replaced with any other method for set membership testing.

\section{Preliminaries}

\paragraph*{Bloom Filters}
\label{BFintro}
The Bloom Filter is an array of $m$ bits which represents a set $S$ of $n$ elements. It is initialized with all bits set to 0. During construction, we apply $\eta$ universal hash \cite{carter1978exact} functions $\{h_1, h_2...h_{\eta}\}$ with range $m$ to the elements of $S$. We set the bits at the respective locations $\{h_1(x), h_2(x)...h_{\eta}(x)\}$ for each key $x \in S$. 
Once the construction is done, the Bloom Filter can be used to determine whether a query $q\in S$ by calculating the AND of the bits at the $\eta$ locations: $h_1(q), h_2(q)...h_{\eta}(q)$. The output will be $\mathrm{True}$ if all the $\eta$ locations are $1$ and it will be $\mathrm{False}$ if at least one location is $0$. 
Bloom Filters have no false-negatives as every key $x \in S$ will set all the bits at locations $\{h_1(x), h_2(x)...h_{\eta}(x)\}$. However, there are false-positives introduced by hash collisions.
The false-positive rate of the Bloom Filter, p, is given by: $  p = \left(1-\left[1-{\frac {1}{m}}\right]^{\eta n}\right)^{\eta} \approx \left(1-e^{-\eta n/m}\right)^{\eta} $.
\noindent To minimize the false-positive rate, we use $\eta = -\frac{\log p}{\log 2}$ and $m = -n \frac{\log p}{\log 2}$.
The size of a Bloom Filter grows linearly in the cardinality of the set it represents. 
\section{RAMBO: Repeated And Merged Bloom Filters}
\label{RAMBO}
\begin{figure}[h]
    \centering
    \begin{minipage}{0.48\textwidth}
    \includegraphics[scale=0.45]{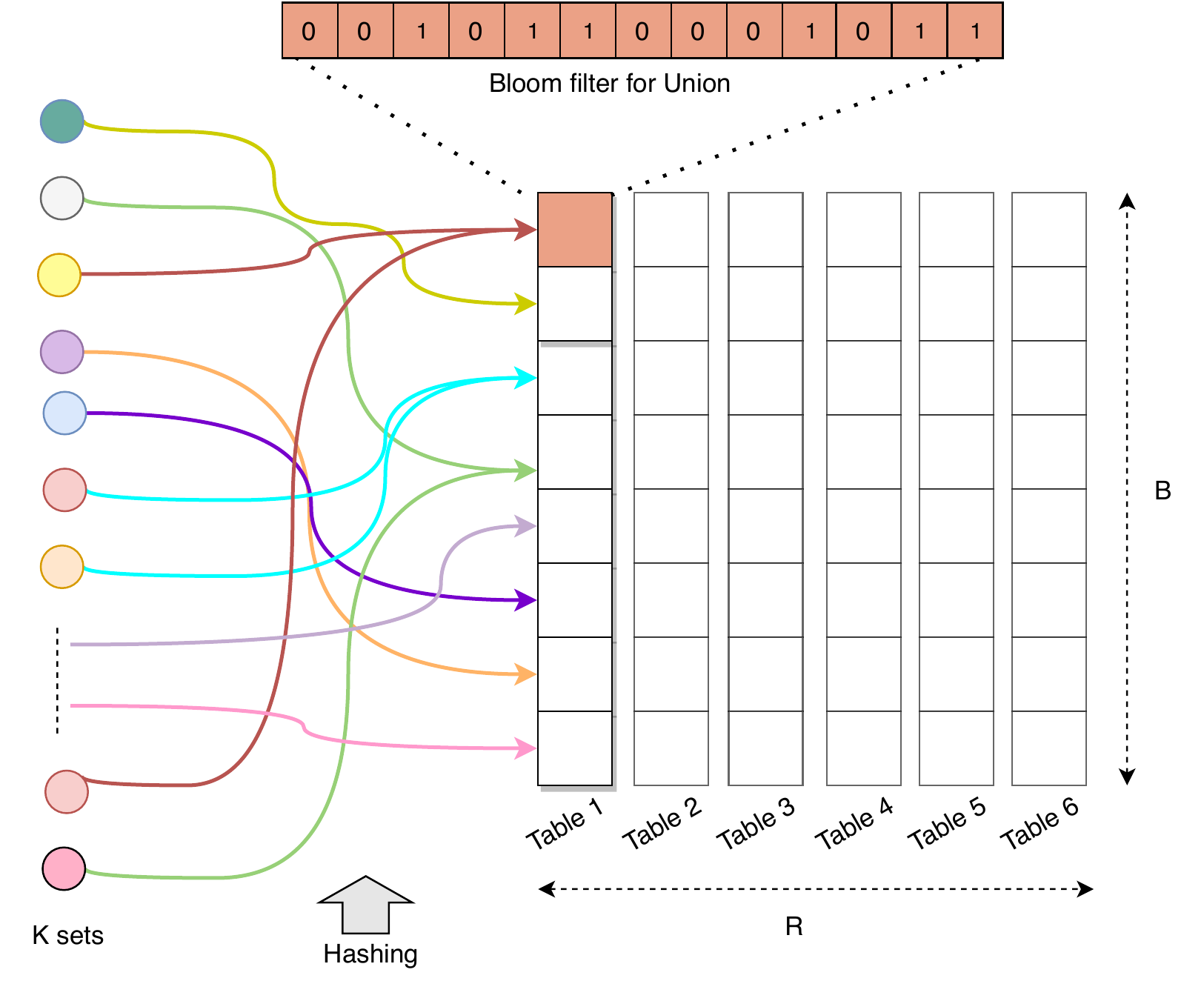}
    \end{minipage}
    \hfill
    \begin{minipage}{0.48\textwidth}
    \begin{algorithm}[H]
    \begin{algorithmic}
    \STATE {\bf Input:} Set $\mathcal{S}$ of $K$ sets 
    \STATE {\bf Result:} RAMBO (size: $B \times R$ )
    \STATE Generate $R$ partition hash functions $\phi_1(\cdot), ... \phi_R(\cdot)$
    \STATE RAMBO $\leftarrow B\times R$ array of Bloom Filters
    \WHILE{Input $S_i$}
        \FOR{term $x \in $ $S_i$}
            \FOR{$d = 1, ... R$}
                \STATE Insert($x$, RAMBO[$ \phi_d(x),d$])
            \ENDFOR
        \ENDFOR 
    \ENDWHILE
     \end{algorithmic}
      \caption{Algorithm for insertion in RAMBO architecture}
      \label{alg:construction}
    \end{algorithm}
    \end{minipage}
    \caption{The figure shows the RAMBO architecture and the insertion process. The construction of first repetition is highlighted. Here the $K$ sets are randomly partitioned (via a 2-universal hash function \cite{carter1978exact}) into cells. Each cell is a Bloom Filter of the union of merged sets. In the architecture there are $R$ such tables containing $B$ bucket each. Refer algorithm on the right for construction.}
    \label{fig:RAMBOarch}
\end{figure}

The architecture of RAMBO (Figure \ref{fig:RAMBOarch}) comprises array of $R$ tables, each containing $B$ cells. 
In each table, the $K$ sets are partitioned into $B$ groups, one for each cell. The cells contain Bloom Filters for the Union (BFU) of the sets assigned to their corresponding partitions. Each BFU is a compressed representation of a group of sets, rather than an individual set representation. The tables in RAMBO are independent, with different hash functions, to help single out a common set that contains the query. The $K$ sets are assigned to one cell in each table and are thus hashed $R$ times. Due to the universality of the hash functions, every cell of a table in RAMBO contains an expected $K/B$ sets from $\mathcal{S}$.

\paragraph{Intuition}
We have $K$ sets partitioned into $B$ groups, where $2 \le B \ll K$. Now, given a query term $q$, if we query each partition, we can determine which partition contains $q$. We refer to this partition as $A1$. Thus, with only $B$ Bloom Filter queries, we have reduced the number of candidate sets from $K$ to $\frac{K}{B}$ in expectation. If we independently repeat this process again, we find another partition $A2$ that contains $q$. Our pool of candidate sets is now the set intersection of $A1$ and $A2$, which in expectation has size $\frac{K}{B^2}$. With more repetitions, we progressively rule out more and more options until we are left with only the sets that contain $q$.

The critical insight is that each repetition reduces the number of candidates by a factor of $\frac{1}{B}$, which decreases \textit{exponentially} with the number of repetitions. RAMBO uses this observation to identify the correct sets using an exponentially smaller number of Bloom Filter queries.

Since RAMBO is an extension of the Count-Min Sketch (CMS) data structure \cite{Cormode2005CMS}, most theoretical guarantees carry forward. We replace the counters in the CMS with Bloom Filters. Instead of adding counters in the CMS, we merge sets of k-gram terms. The querying procedure of the CMS is replaced with an intersection over the merged sets to determine which sets contain a query term.


\subsection{Construction}
We assign sets based on a partition hash function $\phi(.)$ that maps the set identity to one of $B$ cells. We use $R$ independent partition hash functions $\{ \phi_1, \ \phi_2, ..., \phi_R\}$. 
Suppose we want to add a set of terms in DOC-1 to RAMBO. We first use the partition functions $\phi_i($DOC-1$)$ to map DOC-1 in repetition $i$, $\forall \ i \in \{0,\ 1, ..., R \}$. Then we insert all the terms (k-grams) of DOC-1 in $R$ assigned BFUs. Refer Algorithm \ref{alg:construction}. The Bloom Filter insertion process is defined in section \ref{BFintro}. We define the size of each BFU based on the expected number of insertions in it. This is further analyzed in section \ref{paramSelection}.

Our RAMBO data structure is merely a CMS (of size $B\times R$) of Bloom Filters. Clearly, this structure is conducive to updates with a data stream. Every new term from a document is hashed to unique locations in $R$ tables. The size of BFU can be predefined or a scalable Bloom Filter \cite{scalableBF} can be used for adaptive size. 

\subsection{Query}
\label{queryMethod}
\begin{figure}[h]
\begin{minipage}[b]{0.5\textwidth}
    \includegraphics[scale=0.56]{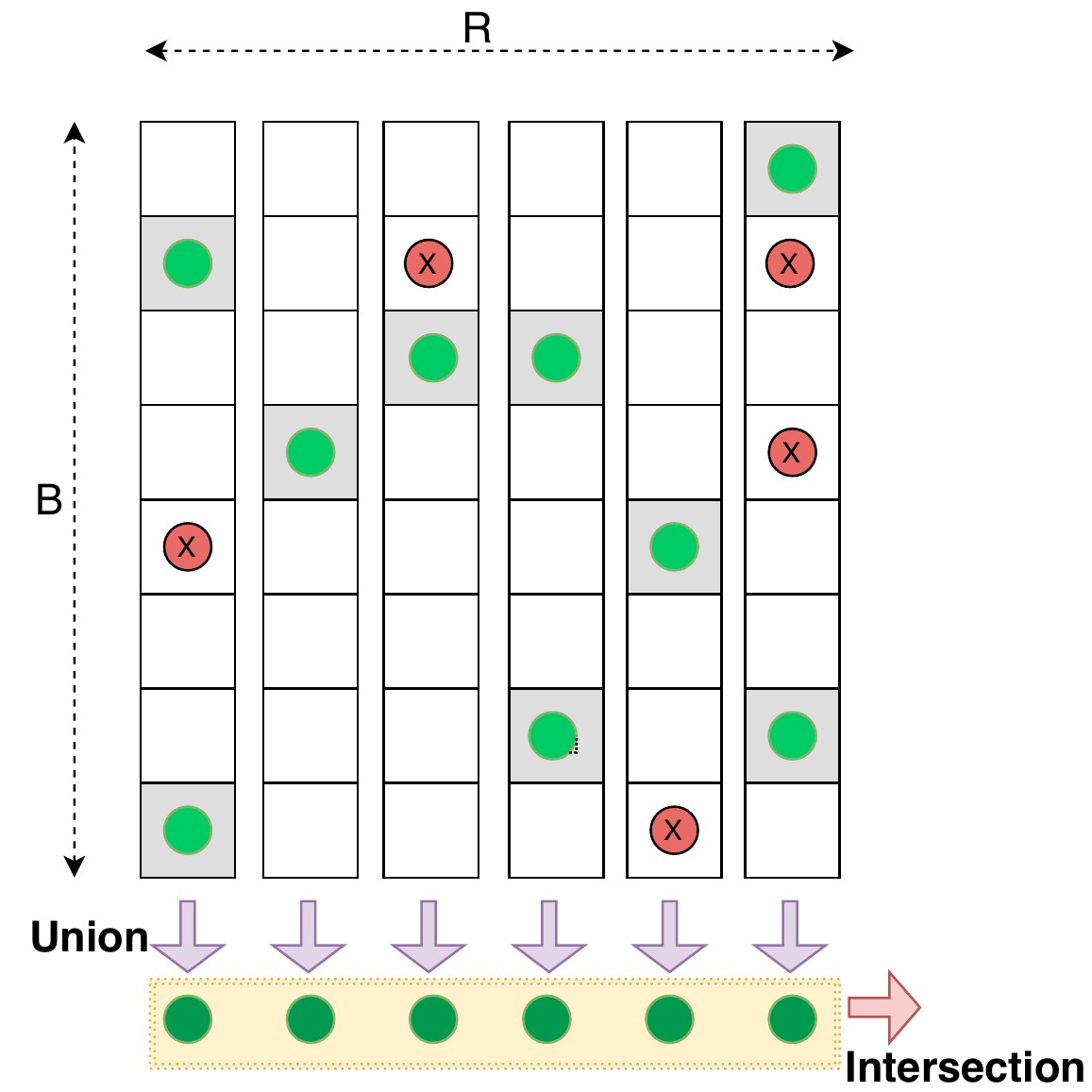}
    \label{fig:RAMBOpath}
\end{minipage}
\hfill
\begin{minipage}[b]{0.5\textwidth}
\centering
\begin{algorithm}[H]
\begin{algorithmic}
\STATE {\bf Input:} query $q \in \Omega$
\STATE {\bf Architecture:} RAMBO (size: $B \times R$ )
\STATE {\bf Result:} $A \subset \mathcal{S} $, where $q \in S_i \ \forall S_i \in A$
\STATE $Q = $Terms$(q)$
\FOR{$r=1:R$}
    \STATE $G_r = \phi$
    \FOR {$b=1:B$}
        \STATE $G_r = G_r \ \cup \ $SetIDs(M$[b,r])\ \textbf{if}\ Q \in$ BF(M[$b,r$]) 
    \ENDFOR
\ENDFOR
\STATE $A = \cap_{i} G_i$ \COMMENT{final returned set ID's} \\

\STATE {\bf Define:} Q\ $\in \ $BF(M$[b,r]$)
  \STATE $\forall x \in $Q$ \ \ \rightarrow x \in \ $BF(M$[b,r]$), Q$ \ \subset \ $M$[b,r]$
 \end{algorithmic}
 \caption{Algorithm for membership testing using RAMBO architecture}
 \label{alg:membershipTest}
\end{algorithm}
\end{minipage}
\caption{
    For a given query each table of RAMBO returns one or more BFUs (represented by dots) where the membership is defined. Here the red dot represents the false-positive and green dot represents true positive. The membership for sets for each table is defined by the union. It is followed by the intersection of the returned sets, which define the final membership of the query $q$.}
\end{figure}

A term can occur in multiple sets due to its multiplicity. The RAMBO architecture shown in figure \ref{fig:RAMBOarch} returns at least one cell per table at query time. Due to the possibility of false-positives, it can return multiple cells per table. This is a slight departure from standard CMS. Our query process, therefore, first takes the union among the returned cells from each table and then the intersection of all those unions across $R$ tables. The union and intersection can be accomplished using fast bitwise operations. Algorithm \ref{alg:membershipTest} summarizes the flow. The extra query time incurred during this process is analyzed in Section \ref{queryTime}. Set $A\subset \mathcal{S}$ as shown in Algorithm \ref{alg:membershipTest} is the final returned set of matched documents. 

\subsubsection{Large Sequence Query}
\label{Fullsequencequery}
To query a larger term sequence with length $n$, we simply use a sliding window of size $k$ to go through the entire sequence. This will create a set of terms $Q$ to query. Then we iterate over terms in $Q$ and membership test each term. The final output should be the conjunction of all returned outputs from each term in $Q$.
Since Bloom Filter does not have any false-negatives, we are guaranteed to obtain a valid result. We only need to perform exponentially less (in the cardinality of $Q$) number of membership tests as the first returned FALSE will be conclusive. It is interesting to note that the result will be governed by the rarest $k$-gram term of the given query sequence.

\section{Analysis}
\label{TheroryAnalysis}
RAMBO is a matrix ($B \times R$) of Bloom Filters which enjoys all the properties of arrays of Bloom Filters, i.e., streaming updates and bitwise operations. RAMBO has two important parameters $R$ and $B$ that control the resource-accuracy trade-off. In this section, we will analyze false-positive rate, query time and index size to find the optimal values of R and B. 

\subsection{False-Positives}
\label{FPanal}
Our first claim is that RAMBO cannot report false-negatives. This follows trivially from the hashing-based merging procedure and the fact that each BFU cannot produce false-negatives \cite{bloom1970space}. Next, we begin by finding the false-positive rate of one set and extend this result to all $K$ sets. 




\begin{lemma} False-positive rate \\
Given the RAMBO data structure with $B\times R$ BFUs, each with false-positive rate $p$ and query $q$, we assume that $q$ belongs to no more than $V$ sets. Under these assumptions, the probability of incorrectly reporting that $q \in S_i$ when $q \not \in S_i$ is 
$$F_p = \left(p \left(1 - \frac{1}{B}\right)^V + 1 - \left(1 - \frac{1}{B}\right)^V\right)^R $$ 
where $p$ is the individual false-positive rate of the BFUs.
\label{fpTheorem}
\end{lemma}
False-positives can be introduced by BFUs by its Bloom Filter property or by violating the separability of sets. The proof is detailed in supplementary material. Using this theorem, we can construct the overall failure rate for the MSMT problem. 


\begin{lemma} RAMBO failure rate \\
Given a RAMBO data structure with $B\times R$ BFUs, each with false-positive rate $p$ and query $q$, we assume that $q$ belongs to no more than $V$ sets. Under this assumption, the probability ($\delta$) of reporting an incorrect membership status for any of the $K$ sets is upper bounded by
$$\delta \leq K \left(1 - (1 - p)\left(1 - \frac{1}{B}\right)^V\right)^R $$ 
where $p$ is the individual false-positive rate of the BFUs.
\label{overallFailureRate}
\end{lemma}

A direct consequence of lemma \ref{overallFailureRate} is that we need sub-linear RAMBO repetitions (logarithmic in $K$) to obtain any overall failure probability $\delta$. We can state that it is sufficient to keep $R \geq {\log K - \log \delta}$. Proof is detailed in supplementary material.

\begin{theorem}
\label{Rbound}
Given a failure rate $\delta$, and a RAMBO data structure with $B\times R$ BFUs, we need 
$$R = O(\log K - \log \delta)$$
where ($\delta$) is probability of reporting an incorrect membership status for any of the $K$ sets.
\end{theorem}


\subsection{Query Time Analysis}
\label{queryTime}
This section demonstrates that RAMBO achieves the expected sub-linear query time. To query the set membership status of an element $x$, we perform $B \times R$ Bloom Filter look-ups followed by union and intersection operations (Section \ref{queryMethod}). The time cost of a BFU look-up is constant. We need to examine the cost of union and intersection operations.

Since each repetition partitions the $K$ sets, the union operations do not require any computational overhead. The set intersections between repetition, however, require $|X_1| + |X_2|$ operations, where $X_1$ is the set of all active sets in the present repetition and $X_2$ is the set of all active sets in the next repetition. Since there are $R$ repetition, the total cost for the intersection is $\sum_{r = 1 }^R |X_r|$. By observing that $\mathbb{E}[|X_r|] = V + Bp$, we obtain the following theorem. 

\begin{lemma} Expected query time\\
Given the RAMBO data structure with $B \times R$ BFUs and a query $q$ that is present in at most $V$ sets, the expected query time
$$\mathbb{E}[q_t] = BR ( \eta )+ \frac{K}{B}(V + Bp)R$$
where $K$ is the number of sets, $p$ is the BFU false-positive rate, and $\eta $ is the number of hash functions used in BFUs ($\eta =2$ in experiments). 
\end{lemma}
We get $B = \sqrt{KV/\eta}$ from here by minimising ($\nabla_{B} (\mathbb{E}[q_t])=0$) over query time. 
Additionally, by safely assuming $p \leq \frac{1}{B}$ and $R$ according to Theorem~\ref{Rbound}, we get an expression for the query time in terms of the overall failure probability and the total number of sets $K$. Our main theorem is a simplified version of this result where we omit the lower order terms. 

\begin{theorem} MSMT query time\\
Given a RAMBO data structure and a query $q$ that is present in at most $V$ sets, RAMBO solves the MSMT problem with probability $1 - \delta$ in expected query time
$$\mathbb{E}[q_t] = O\left(\sqrt{K} \left(\log{K} - \log {\delta}\right) \right)$$
where $K$ is the number of sets, $p$ is the false-positive rate of each BFU, and $V$ is independent of $K$ and $\delta$.
\label{MSMT_qt}
\end{theorem}

\subsection{Memory Analysis}
We provide an average-case analysis under a simplifying assumption to analyze the expected performance of our method. To make the analysis simpler, we assume that every key has fixed multiplicity $V$. This means every item is present in $V$ number of sets. We define the memory of RAMBO as 

\begin{lemma}
\label{expectedMem}
For the proposed RAMBO architecture with size $B \times R$ and data where every key has $V$ number of duplicates, the expected memory requirement is 
$$\mathbb{E}_v(M) = \Gamma \log K \log (1/p)\sum_{S \in \mathcal{S}}|S| \ \ \ \ \ \ \
\text{where} \ \Gamma = \sum^{V}_{v =1}\frac{1}{v}\frac{(B-1)^{V-2v+1}}{B^{V-1}}$$
\end{lemma}
The expectation is defined over variable $v$ which takes values from \{1,2...V\}. The factor $\Gamma <1$. This expression of $\Gamma $ holds if we are hashing document IDs using a universal hash function. If $B=K$, we will have one Bloom Filter per set. In that case $\Gamma=1$.
We will prove the expression of $\Gamma $ and its variation for any $B<K$ and $V>1$ in supplementary material. 


\section{Experiments}
We perform experiments on two large scale datasets: 1) Genomic data and 2) English web documents.
\subsection{Parameter Selection}
\label{paramSelection}
For a fixed BFU with false-positive rate $p$ and $\eta$ number of hashes, the size of BFU depends only on the number of insertions (Section \ref{BFintro}). One way to determine the BFU size is to preprocess the data by counting the terms in all sets \cite{}. Luckily, as we are pooling the sets, the number of insertions for each BFU has less variance.
We calculate the average set cardinality from a tiny fraction of the data at the beginning and use it to set the size for all the BFUs. Hence, we truly achieve the RAMBO insertion on streaming datasets. 
 The parameters $B$ and $R$ are chosen empirically keeping in mind the $O(\sqrt{K})$ and $O(log K)$ variation. 



\subsection{Genomic sequence indexing}
\label{genomicExpe}
\noindent\textbf{Dataset}: We use the 170TB whole genome sequence (WGS) dataset (containing 447833 files) as described in~\cite{b3} and \cite{b12}. It is the set of all bacterial, viral, and parasitic WGS data in the European Nucleotide Archive (ENA) as of December 2016. The terms in each file (set) are created by taking overlapped sliding window of length 31 on the gene sequence comprised of 4 nucleotides (ATGC). Term is called k-mer here.

\noindent\textbf{Baselines}: We compare our methods with COBS (Compact bit-sliced signature index)~\cite{Bingmann2019COBSAC} (Index based on array of Bloom Filters) and the Bloom Filter tree-based methods, SSBT ~\cite{b7} and HowDeSBT ~\cite{HowDeSBT}. To ensure a fair comparison, we have selected baseline hyper-parameters from their papers and kept a common (approximately) false-positive rate. All the baseline implementations and RAMBO are in C++.

\begin{table*}[h]
\fontsize{7.9}{13}\selectfont

  \centering
  \caption{Performance comparison between RAMBO and baselines. 
  For fair comparison, we tried to keep the false-positive rate on equal level for baselines. The results are generated on a set of 1000 queries with exponentially distributed ($\alpha = 100$) term multiplicity $V$ . We could not benchmark HowDeSBT (HowDe) on a larger scale due to memory and time constraints on the cluster we use. It exceeds 192 GB (available RAM) mark. 2nd part of Table shows how the memory and construction time varies with $K$} 
  \begin{tabular}{ |c|c|c|c|c|c|c|c|c|c|c|c|c|c|c|c|c|c|c| } 
 \hline
 & \multicolumn{4}{|c|}{ Time per query (ms) (cpu time)} & \multicolumn{4}{|c|}{ False-positive rate } & \multicolumn{4}{|c|}{ Size} \\
 \hline
\#files &  HowDe   &  SSBT   &   COBS &  RAMBO  &   HowDe    &   SSBT  &   COBS   &    RAMBO &   HowDe      &  SSBT   &   COBS  &   RAMBO \\
\hline 
100 &  5.52  & 10.673 & 0.18& {\bf 0.014} & 0.00546  & 0.00845 & 0.008 & 0.011  & 25GB  & 4.2G & {\bf 1.6G} & 3.5GB\\ 
\hline 
200 & 10.72   & 35.953& 0.39 & {\bf 0.017}& 0.00565 & 0.00826& 0.008 & 0.01 & 49GB  & 8.5G & 7.0G & {\bf 6.3GB}\\ 
\hline 
500 & 23.94   & 72.289 & 1.07 & {\bf 0.04} & 0.00581  & 0.00878& 0.008 & 0.0093 & 152G  & 21G & {\bf 7.9G} & 13.9GB\\ 
\hline
1000 & -  & 143.27 & 1.76 & {\bf 0.07}  &-  & 0.00834 & 0.008 & 0.010 & -   & 42GB & {\bf 16G}& 23.2GB\\
\hline
2000 & -  & 259.127 & 2.66 & {\bf 0.10} &-  & 0.00815& 0.008 &  0.0129 & -  & 83GB & {\bf 21G}& 35 GB\\
\hline
\end{tabular}
  
\label{HowDeSBTtable}
\end{table*}
\vspace{-0.2cm}
\begin{table*}[h]
\fontsize{7.9}{13}\selectfont
  \centering
  \begin{tabular}{|c|c|c|c|c|c|c|} 
 \hline
\# of files & 100  &200& 500 & 1000 & 2000\\
\hline 
\hline 
RAMBO Construction time & 7.2 min & 14.4 min & 37 min & 102 min & 234 min\\ 
\hline 
Number of partitions (B)  & 15 & 27 & 60 & 100 & 150 \\ 

\hline
\end{tabular}
  
\label{HowDeSBTtable2}
\end{table*}

\noindent\textbf{Parameters}: For HowDeSBT, the Bloom Filter size is $2 \times 10^9$. HowDeSBT only supports $1$ hash function. 
For SSBT, we use $4$ hash functions and set Bloom Filter size to $5 \times 10^8$ bits.
For COBS, we use $3$ hash functions and set false-positive rate to 0.008.
For RAMBO, we set the number of repetitions $R$ as 2 and $B$ as $15, 27, 60, 100$ and $150$ for number of set insertions $100, 200, 500, 1000$ and $2000$. The Bloom Filter size is kept to $10^9$ bits. For fair comparison, we tried to keep the false-positive rate on equivalent level.

\noindent\textbf{Evaluation Metrics}: 
 Creating a test set with ground truth requires a very time-consuming procedure of generating inverted indices. Therefore, we calculated the false-positive rate by creating a test set of $1000$ randomly generated $30$ length k-mer terms.
 Length $30$ ensures that there is no collision from the existing k-mers in the RAMBO data structure. These k-mers were assigned to $V$ files (distributed exponentially with $alpha=100$) randomly. Being much smaller than the actual size of the dataset, this test data makes an insignificant change in the load capacity of RAMBO Bloom Filters. 

 \textbf{System and Platform Details}: We ran the experiment on a cluster with multiple $40$ core Intel(R) Xeon(R) Gold 6230 CPU @ 2.10GHz processor nodes. Each node has 192 GB of RAM and 960 GB disk space. All the experiments apart from RAMBO construction on full dataset, are performed on a single node. Multi-threading is not used for querying.
 
  From Table \ref{HowDeSBTtable} we can see that RAMBO's is much faster (from around \textbf{25x} to \textbf{2000x}) in query time than the baselines. Furthermore RAMBO achieves a \textbf{very small index size} (close to theoretical lower bound of array of Bloom Filters).
 Repetitions in RAMBO decrease the false-positive rate exponentially, making it at par with the baselines.
 
\paragraph{Smart parallelism- Indexing full 170TB WGS dataset in 14 hours from scratch}

For this sheer amount of data, we parallelize the computation by partitioning the RAMBO data-structure over 100 nodes. Each node contains a small RAMBO data-structure indexing $1/100$ of the whole dataset, which is $4605$ files in our case. In the streaming setting a file (set of terms) is routed to a BFU of a node randomly. We achieve this by using a two-level hash function.
\begin{equation}
    (b \times \tau(D_j)) + \phi_i(D_j) 
\end{equation}

The universal hash function $ \tau (.)$ assigns file to a random machines and then the independent node-local universal hash function $\phi_i(.)$ assigns the file to the local BFU. This eliminates costly transmission of data among the nodes. The data-structure on each node has size $B =500$ and $R=5$. Stacking them vertically makes the complete RAMBO data-structure of size $B =50000$ and $R=5$. 
This process preserves the randomness of set insertion,  i.e., the probability of any two sets colliding is exactly $1/B$, where $B$ is the total range ($50000$ in this case). 

 \begin{figure}[!htbp]
    \includegraphics[scale=0.5]{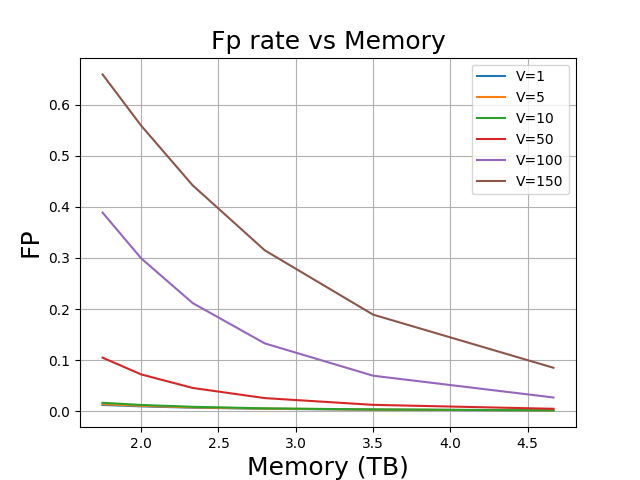}
    \qquad
    \fontsize{12}{16}\selectfont
    \begin{tabular}[b]{|c|c|c|}\hline
      & Query Time   &  Index     \\
          & (cpu time in ms) &  size\\ 
  \hline 
  Fold 2 & 66.5 & 7.13 TB  \\  
 \hline
  Fold 4 & 43.5  & 3.6 TB  \\  
 \hline 
  Fold 8 & 26.25  & 1.78 TB \\  
 \hline 

    \end{tabular}
    \captionlistentry[table]{A table beside a figure}
    \caption{Figure shows the false-positive rate of RAMBO for different values of V (k-mer multiplicity per $4605$ sets) and memory. 
It is very interesting to note that the false-positive rates are very low if querying a rare k-mer. For a full sequence search, the returned result depends solely on the rarest k-mer term. Hence our method returns very accurate (low false-positives) results. The table shows the cpu time (in ms) per query of the k-mer averaged over 1000 queries. Each column shows the different number of RAMBO folds. Second row shows the memory size (in TB) of RAMBO for each fold.}
    \label{fig:6}
  \end{figure}

This interesting parallel insertion trick results in fully constructed RAMBO only in  \textbf{14 hours (8 hours to download and 6 hrs to insert)}. It is the round off approximate time of the most time taking jobs. Here we have to ensure that all machines use the same parameters ($B, R$, Bloom Filter size and hash function $\tau(.)$, $\phi(.)$ and $h(.)$) as well as the random seeds. The consistency of seeds across machines and the larger than required $B$ and $R$ allow us to flexibly reduce the size of RAMBO later by doing bitwise OR between the corresponding BFUs of first half of RAMBO over the other half(vertically). Each of this process reduces the index size $B \times R$ to $\frac{B}{2} \times R$. This is called folding over. Refer Figure \ref{fig:6}. More details can be found in supplementary material.

\subsection{Document indexing}
\noindent\textbf{Datasets}: We use a sample from Wiki dump \cite{wiki-dump} and the popular TREC Category B ClueWeb09 dataset\cite{clueweb}. The Wiki dump sample has 17618 documents curated by the authors of BitFunnel \cite{BitFunnel}. The ClueWeb09 dataset has 28 Million (non-spam) documents of English language. Both the datasets were pre-processed by removing stop words, keeping only alpha-numeric, and tokenizing as word unigrams. Wiki dump is $207$ MB and Clueweb is $60GB$ after pre-processing.

\noindent\textbf{Parameters}: BitFunnel \cite{BitFunnel} is indexed with $0.1$ bit density with optimal treatment. For RAMBO, we set $B=2000, R =2$, and the size of each BFU to $200000$ bits in the wiki-dump index. For ClueWeb09, we choose $B=2 \times 10^4, R =4$, and the size of each BFU to be $2 \times 10^6$ bits. 

\noindent\textbf{Baseline}: We compare with BitFunnel \cite{BitFunnel} on wiki-dump. BitFunnel is popular for search indexing (used in Microsoft Bing). 

\noindent\textbf{Evaluation Metric}: In the Wiki dump experiment, the query set of single words is taken from the TREC terabyte track 2005 \cite{TREC} efficiency topics. We made an inverted index for ground-truth as it is possible to do on $17K$ documents in a reasonable time. For ClueWeb09, we created a query set of randomly generated terms other than what is present in the data and inserted them using an exponentially distributed term multiplicity $V$, similar to the experiment on genomic data.
Experiments were performed on the same system as in section \ref{genomicExpe}. Query is performed sequentially on single core and thread for fair comparison. Refer Table \ref{BitFunnelTable}.

The code to replicate all the experiments is available \footnote{https://github.com/gaurav16gupta/RAMBO_MSMT}.

\begin{table*}[h]
\fontsize{10}{16}\selectfont
  \centering
  \caption{ Query time is time per query (cpu time) in milisec. We can see that the RAMBO takes \textbf{4x less memory} than the Bitfunnel index and is much \textbf{faster in index creation }. Furthermore, RAMBO can execute the queries in \textbf{3/4} of the time Bitfunnel takes. Motivated by this we also indexed ClueWeb09 dataset on RAMBO. } 
  \begin{tabular}{ |c|c|c|c|c|c|c|c|c|c|c|c|c|c|c|c|c|c|c|c|c| } 
 \hline
 & \multicolumn{2}{|c|}{ Query time (ms)} & \multicolumn{2}{|c|}{ False-positive rate } & \multicolumn{2}{|c|}{ Size} 
 & \multicolumn{2}{|c|}{ Construction time } \\
 \hline
Dataset &  Bitfunnel  &  RAMBO &  Bitfunnel  &  RAMBO &  Bitfunnel  &  RAMBO & BitFunnel  &  RAMBO     \\

\hline
Wiki-dump & 0.12 & \textbf{0.09} & 0.0347 & 0.17 & 197.9 MB & \textbf{50 MB} & 43.2sec & \textbf{1.75sec} \\
\hline 
\end{tabular}
\label{BitFunnelTable}
\end{table*}

\begin{table*}[h]
\fontsize{10}{16}\selectfont
  \centering
  \begin{tabular}{ |c|c|c|c|c|c|c|c|c|c|c|c|c| } 
 \hline
Dataset &   Query time (ms) &   False-positive rate  &   Size &  Construction time    \\
\hline 
ClueWeb09  &  170 & 0.116068  & 19 GB & 144min \\ 
\hline 
\end{tabular}
  \end{table*}

\section{Discussion}
RAMBO provides a solid trade-off between false-positive rate and query time while retaining all desirable properties of Bloom Filter and the bitsliced data structure. Due to cheap updates, RAMBO takes very little time for index creation. Table \ref{BitFunnelTable} shows almost $25\times$ reduction in construction time. RAMBO performs update on stream and is embarrassingly parallel for both insertion and query. RAMBO enjoys zero false-negative rate, low false-positive rate and fast query time. The false-positive rate of RAMBO is very low for low term multiplicity (Figure \ref{fig:6}). This low false-positive rate is guaranteed for full sequence/phrase query, as the rarest of the term dominates. Therefore, RAMBO can perform quick and accurate check of an unknown and rare gene-sequence! Furthermore, due to sub-linear scaling, RAMBO becomes more efficient in memory at a large scale when compared to existing methods. This property will allow RAMBO to be used as an efficient search engine for extreme-scale applications. 

\newpage
\section*{Broader Impact}

There are many real world applications where RAMBO could prove useful. 

1) It could be used in an effort to fill genomic databases with rare or unclassified sequences. If a query sequence is contained in few or no files, it could then be determined to be a valuable sequence worth uploading to public repositories. Also, RAMBO can serve as a fast and accurate tool to trace the origin of a newly introduced gene signature. 

2) In applications like cancer genomics and antibiotic-resistant pathogen sequencing, the sequences of interest will very likely be sequences containing at least one or more structural variants. Again, these variants which lead to the presence of rare $k$-mers will be handled well by RAMBO. 

3) In the area of pan genomics, RAMBO could be utilized to verify whether newly sequenced variants should be incorporated into a linear or graph pan genome. This could aid in efficiently cataloguing genetic diversity and variations across the human population.

4) RAMBO brings a new paradigm of Bloom Filter based search engine that can be used as a fast and memory efficient index of extreme-scale database. Any organisation that needs a search utility can potentially benefit from RAMBO.



\bibliographystyle{plain}
\bibliography{template}

\begin{thebibliography}{10}

\bibitem{wiki-dump}
{Sample wikipedia corpus }.
\newblock Bitfunnel,
  \url{http://bitfunnel.org/wikipedia-as-test-corpus-for-bitfunnel}.

\bibitem{clueweb}
{The ClueWeb09 Dataset}.
\newblock The Lemur Project, \url{https://www.lemurproject.org/clueweb09.php/}.

\bibitem{b12}
{The European Bioinformatics Institute (EBI): European Nucleotide Archive (ENA)
  Resource}.
\newblock The European Bioinformatics Institute (EBI) FTP Site,
  \url{http://ftp.ebi.ac.uk/pub/software/bigsi/nat_biotech_2018/ctx/}.

\bibitem{TREC}
{TREC terabyte track 2005}.
\newblock TREC, \url{https://trec.nist.gov/data/terabyte05.html}.

\bibitem{scalableBF}
Paulo~S{\'e}rgio Almeida, Carlos Baquero, Nuno Pregui{\c{c}}a, and David
  Hutchison.
\newblock Scalable bloom filters.
\newblock {\em Information Processing Letters}, 101(6):255--261, 2007.

\bibitem{Bingmann2019COBSAC}
Timo Bingmann, Phelim Bradley, Florian Gauger, and Zamin Iqbal.
\newblock Cobs: a compact bit-sliced signature index.
\newblock In {\em SPIRE}, 2019.

\bibitem{bloom1970space}
Burton~H Bloom.
\newblock Space/time trade-offs in hash coding with allowable errors.
\newblock {\em Communications of the ACM}, 13(7):422--426, 1970.

\bibitem{simpleMPH}
Fabiano~C Botelho, Rasmus Pagh, and Nivio Ziviani.
\newblock Simple and space-efficient minimal perfect hash functions.
\newblock In {\em Workshop on Algorithms and Data Structures}, pages 139--150.
  Springer, 2007.

\bibitem{b3}
Phelim Bradley, Henk~C den Bakker, Eduardo~PC Rocha, Gil McVean, and Zamin
  Iqbal.
\newblock Ultrafast search of all deposited bacterial and viral genomic data.
\newblock {\em Nature biotechnology}, 37(2):152, 2019.

\bibitem{networkapplications}
Andrei Broder, Michael Mitzenmacher, and Andrei Broder I~Michael Mitzenmacher.
\newblock Network applications of bloom filters: A survey.
\newblock In {\em Internet Mathematics}, pages 636--646, 2002.

\bibitem{carter1978exact}
Larry Carter, Robert Floyd, John Gill, George Markowsky, and Mark Wegman.
\newblock Exact and approximate membership testers.
\newblock In {\em Proceedings of the tenth annual ACM symposium on Theory of
  computing}, pages 59--65. ACM, 1978.

\bibitem{cohen2003spectral}
Saar Cohen and Yossi Matias.
\newblock Spectral bloom filters.
\newblock In {\em Proceedings of the 2003 ACM SIGMOD international conference
  on Management of data}, pages 241--252. ACM, 2003.

\bibitem{Cormode2005CMS}
Graham Cormode and Shan Muthukrishnan.
\newblock An improved data stream summary: the count-min sketch and its
  applications.
\newblock {\em Journal of Algorithms}, 55(1):58--75, 2005.

\bibitem{crainiceanu2013bloofi}
Adina Crainiceanu.
\newblock Bloofi: a hierarchical bloom filter index with applications to
  distributed data provenance.
\newblock In {\em Proceedings of the 2nd International Workshop on Cloud
  Intelligence}, page~4. ACM, 2013.

\bibitem{BitFunnel}
Bob Goodwin, Michael Hopcroft, Dan Luu, Alex Clemmer, Mihaela Curmei, Sameh
  Elnikety, and Yuxiong He.
\newblock Bitfunnel: Revisiting signatures for search.
\newblock In {\em Proceedings of the 40th International ACM SIGIR Conference on
  Research and Development in Information Retrieval}, pages 605--614, 2017.

\bibitem{HowDeSBT}
Robert~S Harris and Paul Medvedev.
\newblock {Improved representation of sequence bloom trees}.
\newblock {\em Bioinformatics}, 08 2019.

\bibitem{kodama2011sequence}
Yuichi Kodama, Martin Shumway, and Rasko Leinonen.
\newblock The sequence read archive: explosive growth of sequencing data.
\newblock {\em Nucleic acids research}, 40(D1):D54--D56, 2011.

\bibitem{mitzenmacher2002compressed}
Michael Mitzenmacher.
\newblock Compressed bloom filters.
\newblock {\em IEEE/ACM Transactions on Networking (TON)}, 10(5):604--612,
  2002.

\bibitem{b2}
Brad Solomon and Carl Kingsford.
\newblock Fast search of thousands of short-read sequencing experiments.
\newblock {\em Nature biotechnology}, 34(3):300, 2016.

\bibitem{b7}
Brad Solomon and Carl Kingsford.
\newblock Improved search of large transcriptomic sequencing databases using
  split sequence bloom trees.
\newblock In {\em International Conference on Research in Computational
  Molecular Biology}, pages 257--271. Springer, 2017.

\end{thebibliography}

\newpage
\section{Appendix}
\label{supplimentary}
\subsection{Proofs of the Theorems}

\textbf{Lemma 4.1} \textit{ False-positive rate \\
Given the RAMBO data structure with $B\times R$ BFUs, each with false-positive rate $p$ and query $q$, we assume that $q$ belongs to no more than $V$ sets. Under these assumptions, the probability of incorrectly reporting that $q \in S_i$ when $q \not \in S_i$ is 
$$F_p = \left(p \left(1 - \frac{1}{B}\right)^V + 1 - \left(1 - \frac{1}{B}\right)^V\right)^R $$ 
where $p$ is the individual false-positive rate of the BFUs.}

\textbf{Proof}:
The probability of selecting a Bloom filter which should return false is 
 $$\left(1 - \frac{1}{B}\right)$$ if the multiplicity of the key is 1. \\
 If the multiplicity is $V$ then this probability becomes $$\left(1 - \frac{1}{B}\right)^V$$
False-positives can be introduced by Bloom filters property. Hence the probability of returning a false-positive in this case is
 $$ p\left(1 - \frac{1}{B}\right)^V$$
 False-positives can also be introduced by violating the separability condition. Here the correct Bloom filter will return true but it always occurs with an incorrect Bloom filter. The probability is
 $$1 - \left(1 - \frac{1}{B}\right)^V$$
 Therefore, the total false-positive rate for one repetition is
 $$p\left(1 - \frac{1}{B}\right)^V + 1 - \left(1 - \frac{1}{B}\right)^V$$
 \\
 and the total false-positive rate for $R$ independent repetition is
$$F_p = \left( p\left(1 - \frac{1}{B}\right)^V + 1 - \left(1 - \frac{1}{B}\right)^V \right)^R $$ 

\textbf{Lemma 4.2} \textit{RAMBO failure rate \\
Given a RAMBO data structure with $B\times R$ BFUs, each with false-positive rate $p$ and query $q$, we assume that $q$ belongs to no more than $V$ sets. Under this assumption, the probability ($\delta$) of reporting an incorrect membership status for any of the $K$ sets is upper bounded by
$$\delta \leq K \left(1 - (1 - p)\left(1 - \frac{1}{B}\right)^V\right)^R $$ 
where $p$ is the individual false-positive rate of the BFUs.}

\textbf{Proof}: 
 We define failure as being incorrect about at least one of the sets. By the probability union bound, we can set the individual false-positive rate such that we have a bounded overall failure rate.
 
$$\delta \leq K \left( p\left(1 - \frac{1}{B}\right)^V + 1 - \left(1 - \frac{1}{B}\right)^V \right)^R $$ 
$$\delta \leq K \left(1 - (1 - p)\left(1 - \frac{1}{B}\right)^V\right)^R $$

\textbf{Lemma 4.6}
\textit{For the proposed RAMBO architecture with size $B \times R$ and data where every key has $V$ number of duplicates, the expected memory requirement is 
$$\mathbb{E}_v(M) = \Gamma \log K \log (1/p)\sum_{S \in \mathcal{S}}|S| \ \ \ \ \ \ \
\text{where} \ \Gamma = \sum^{V}_{v =1}\frac{1}{v}\frac{(B-1)^{V-2v+1}}{B^{V-1}}$$
}


\textbf{Proof}: 
With $B=1$, the size of the union is $\frac{N}{V}$ where $N = \sum_{S \in \mathcal{S}}{|S|}$ are the total number of insertions and $S \in \mathcal{S}$ is a set.
If we divide these sets into $B$ groups (by using a universal hashing method to hash sets into bins), we will get the union of random groups of sets with every key in a bin has varying number of duplicates $v$ where $v \in \{0,1,2...V\}$. 0 duplicate means the element does not exists and 1 duplicate means the element has only one copy. 
This implies that $v$ is a random variable with some distribution. We are going to derive this distribution.\\
The size of a bucket $b$ is given by :\\
$$|b| = \mathbb {E}\Big[\sum_{i}^{N1}\frac{1}{v}\Big]$$
where there are N1 number of insertions in $b$ bucket and $\frac{1}{v}$ is a random variable, $\frac{1}{v} \in\{{1, \frac{1}{2}, \frac{1}{3}, .....\frac{1}{V}}\}$. 
By linearity of expectation, we can state that
$$|b| = \sum_{i}^{N1}\mathbb {E}\Big[\frac{1}{v}\Big] = \sum_{i}^{N1}\sum^{V}_{v =1}\frac{1}{v} \times p_v$$
We can view $\frac{1}{v}$ as a multiple count reduction factor, which serves as a normalizer for the multiplicity of the keys. 
After randomly dividing the sets into cells, we will analyze the probability ($p_v$) of having $v$ number of duplicates in a bucket of a hash table of size $B$. 
\\
An element can have at most $V$ duplicates or it can have no presence in the bucket. This problem is similar to putting $V$ balls into $B$ bins. The probability of having all $V$ balls in one bucket is given by $\frac{1}{B^{V-1}}$. \\
The probability of having all $V-1$ balls in one bucket and remaining one in any other bin in given by $\frac{1}{B^{V-2}} \times  \frac{B-1}{B}$
\\ 
Hence the probability of getting $v$ balls in one bucket and $V-v$ in remaining others is given by

$$P_v =  \frac{1}{B^{v-1}} \times \Big(\frac{B-1}{B}\Big)^{V-v} $$

From the expression of $|b|$ and $P_v$ and the fact that $\frac{1}{v} \in\{{0, 1, \frac{1}{2}, \frac{1}{3}, .....\frac{1}{V}}\}$,
we have the expected size of a bucket
$$
    |b| = \sum_{i}^{N1}\sum^{V}_{v =1}\frac{1}{v} \frac{1}{B^{v-1}}  \Big(\frac{B-1}{B}\Big)^{V-v}  = \sum_{i}^{N1}\sum^{V}_{v =1}\frac{1}{v}\frac{(B-1)^{V-v}}{B^{V-1}}
$$
 




The expected size of all the cells in all tables is give by
$$ \sum_{i}^{N}\sum^{V}_{v =1}\frac{1}{v}\frac{(B-1)^{V-2v+1}}{B^{V-1}} = \sum_{S \in \mathcal{S}}|S| \sum^{V}_{v =1}\frac{1}{v}\frac{(B-1)^{V-2v+1}}{B^{V-1}} $$
As $N =\sum_{S \in \mathcal{S}}|S|$,
let's set the extra multiplicative factor $\Gamma$ as
$$ \Gamma = \sum^{V}_{v =1}\frac{1}{v}\frac{(B-1)^{V-2v+1}}{B^{V-1}} $$

\begin{figure}[h]
    \centering
    \includegraphics[scale=0.6]{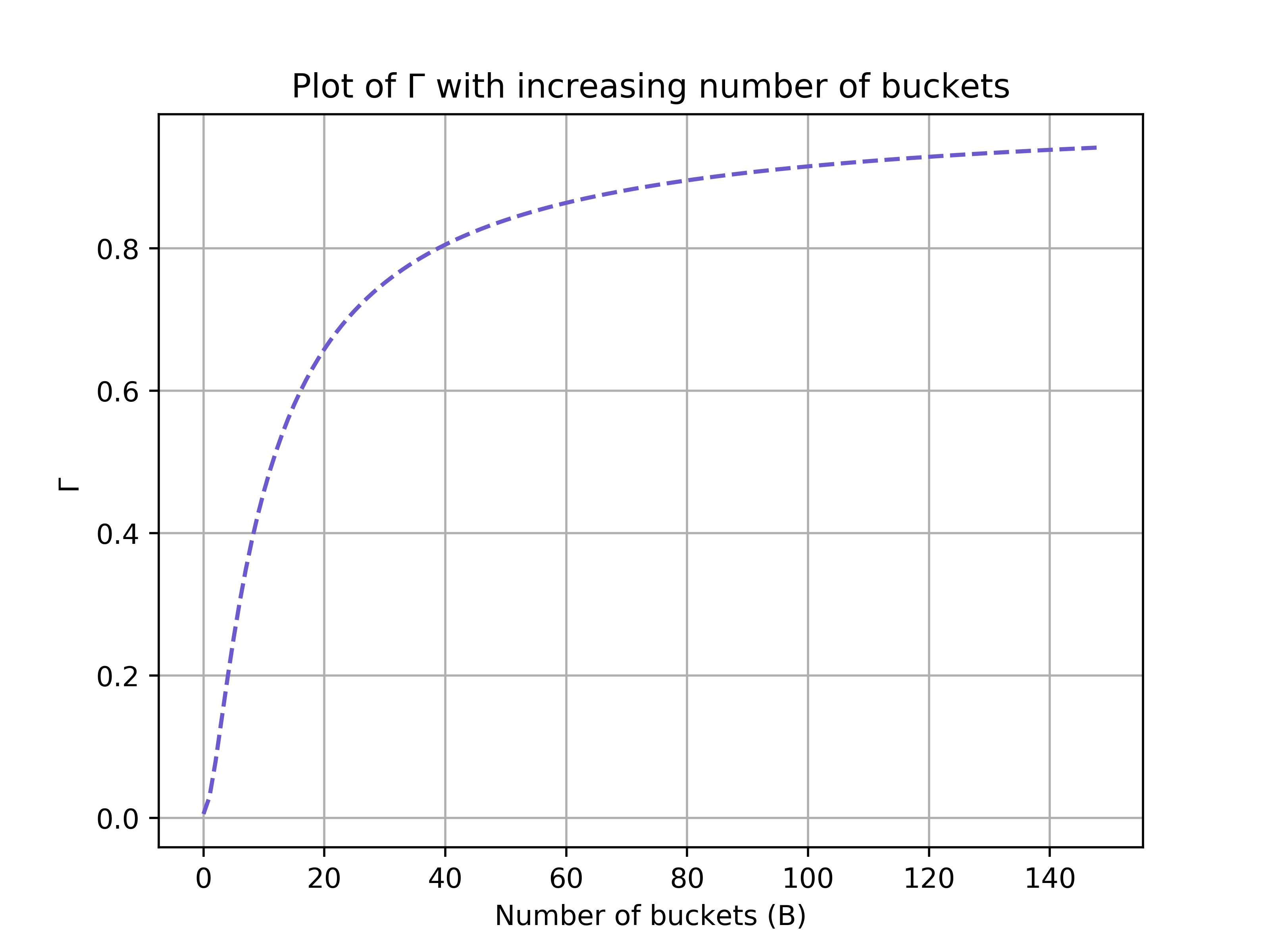}
    \caption{Plot of $\Gamma$ for $V=10$ and $K = 1000$. The factor is less than value 1 for $B<K$ as there are $V>1$ duplicates of the keys. This factor gives the idea of memory saving for one table. }
    \label{fig:MemTH2}
\end{figure}

The derived expression of $\Gamma$ comes from the fact that it is the expected value of the variable $\frac{1}{v}$. As the value taken by this variable lies in $\{1, \frac{1}{2}, \frac{1}{3}, .....\frac{1}{V}\}$ and the probabilities $p_v>0\  \forall v \in [V]$, we can state that the 
$$\frac{1}{v}< \mathbb {E}\Big[\frac{1}{v}\Big] <1 $$
when $B<K$ and $V>1$. Note $\Gamma < 1$. We show the relationship between $\Gamma$ and the number of buckets $B$ in Figure \ref{fig:MemTH2}.\\

\subsection{Experiment: Smart parallelism- Indexing full 170TB WGS dataset in 14 hours from scratch}
\begin{figure*}[h]
\centering
\includegraphics[scale = 0.5]{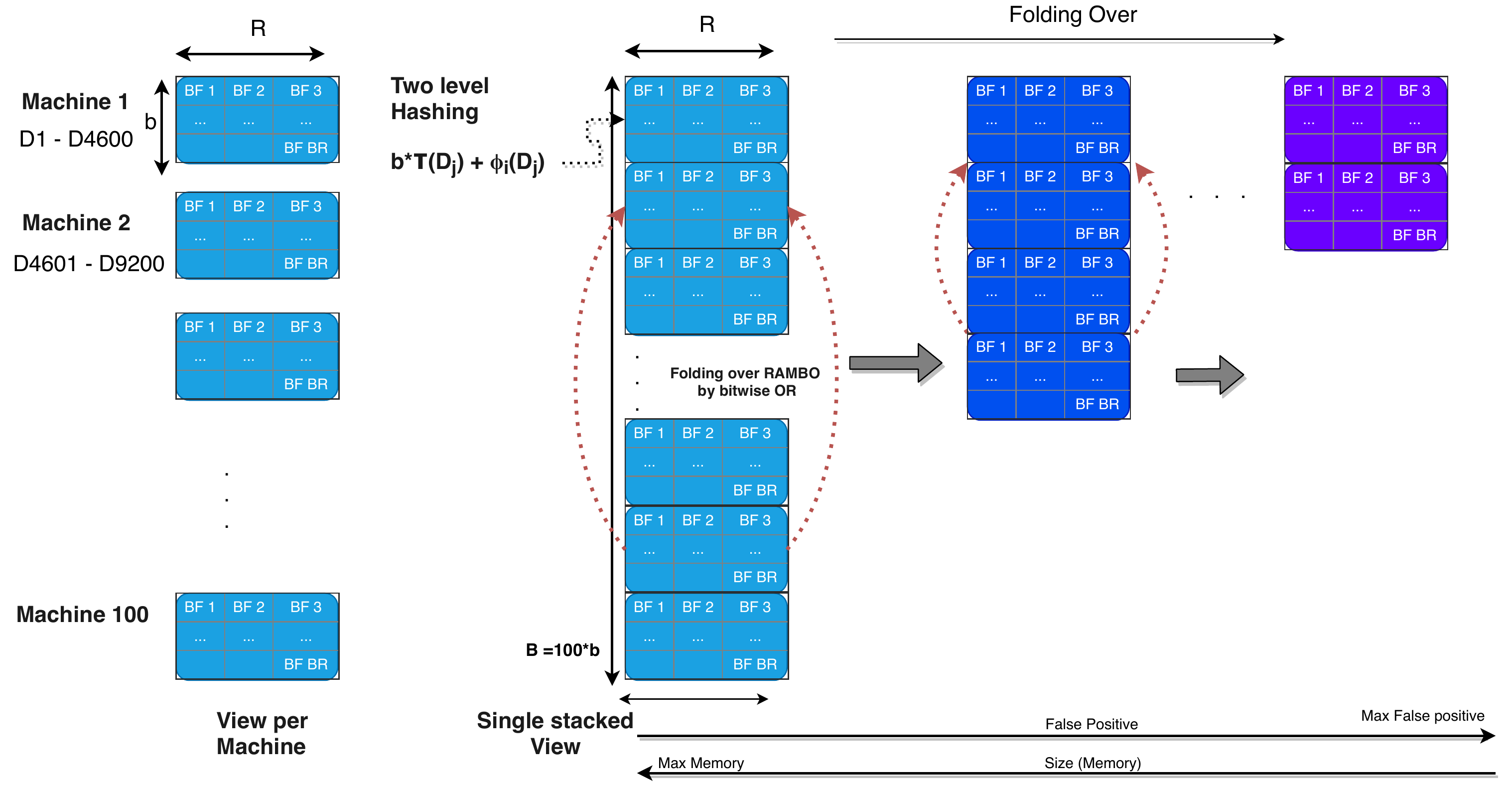}
\caption{This figure illustrates the indexing process of 460500 datasets over a cluster of 100 nodes. Each machine carries the part of RAMBO with size $500 \times 5$ Bloom filters. The dataset is routed to machine via $\tau(.)$ hash function followed by insertion using $\phi_i(.)$. The combined direct routing is done by a two-level hash function equivalent ($b*\tau(.) + \phi(.)$). The single stacked view of RAMBO shows the folding process. The folding is done such that number of repetition $R$ remains the same but $B$ halves, so as the total size. Folding reduces memory progressively by factors of 2, 4, 8... and increases false-positive rate exponentially.}
\label{fig:foldingfig}
\end{figure*}

We first use a random hash function $ \tau (.)$ to assign files to machines and then use an independent smaller machine-local 2-universal hash function $\phi_i(.)$ to assign the file to the local Bloom Filter. It is not difficult to show that this process preserves all the mathematical properties and randomness in RAMBO as the final mapping is again 2-universal, i.e., the probability of any two datasets colliding is exactly $1/B$, where $B$ is the total range (number of partitions in RAMBO). \\
The two-level hash function is given by
\begin{equation}
    (b \times \tau(D_j)) + \phi_i(D_j) 
\end{equation}
For a repetition $i$, where $i \in \{0..R\}$, $b$ is the number of partitions in RAMBO on a single machine and also the range of $\phi(.)$, $D_j$ is the name ID of $j^{th}$ dataset, and the range of $\tau(.)$ is $\{0..100\}$ in our case. Note that this two-level hash function allow us to divide the insertion process into multiple disjoint parts (100 in our case) without repeating any installation of datasets and inter-node communications.


Effectively, each machine will contain a set of $4605$ files. This interesting parallel insertion trick results in 100 independent small data structures in only \textbf{14 hours (8 hours to download and 6 hrs to insert)}. It is the round off approximate time of the most time taking jobs. We have to ensure that all machines use the same parameters ($B, R$, Bloom Filter size and hash function $\tau(.)$, $\phi(.)$ and $h(.)$) as well as the random seeds. The consistency of seeds across machines allows us to flexibly reduce the size of RAMBO later by doing bitwise OR between the two data structures, which we call \emph{folding over}. 
\\ \\ 
\textbf{Folding Over}: The data structures on every machine are independent and disjoint but they have the same parameters and uses the same hash seeds. Since RAMBO is all made of Bloom Filters, we can use a unique folding over property of Bloom Filter that reduces the range of RAMBO from $B$ to $\frac{B}{2}$. The folding over is simply the bit-wise OR between the first half of RAMBO over the other half. This operation is depicted in Figure~\ref{fig:foldingfig}. With this folding over, a one time processing allows us to create several versions of RAMBO with varying memories. 
\\ \\

\end{document}